\documentclass[fleqn,10pt]{wlscirep}
\usepackage{epstopdf} 
\usepackage{bm}
\usepackage{eurosym}
\usepackage{amsmath}
\usepackage{graphicx}
\usepackage{dcolumn}
\usepackage{amssymb}
\usepackage{multirow}
\usepackage{hyperref}
\usepackage{subfigure}
\usepackage{dsfont}
\usepackage{epstopdf}
\usepackage{color}
\usepackage{capt-of} 
\title{Realizing quantum advantage without entanglement in single-photon states}
\author[1,2,*]{A. Maldonado-Trapp}
\author[1]{Pablo Solano}
\author[3]{Anzi Hu}
\author[1]{Charles W. Clark}
\affil[1]{Joint Quantum Institute, National Institute of Standards and Technology and University of Maryland, College Park, MD 20742}
\affil[2]{Departamento de F\'{\i}sica, Universidad de Concepci\'{o}n,
	160-C Concepci\'{o}n, Chile}
\affil[3]{Department of Physics, American University, Washington, DC
	20016, USA}
\affil[*]{alemt@umd.edu}

\begin{abstract}
	We discuss the realization of quantum advantage in a system without quantum entanglement but with non-zero quantum discord. We propose an optical realization of symmetric two-qubit $X$-states with controllable anti-diagonal elements.
This approach does not require initially entangled states, and it can generate states that have quantum discord, with or without entanglement. We discuss how quantum advantage can be attained in the context of a two-qubit game. We show that when entanglement is not present, the maximum quantum advantage is 1/3 bit. A comparable quantum advantage, 0.311 bit, can be realized with a  simplified transaction protocol involving one vs. the three unitary operations needed for the maximum advantage.   
\end{abstract}
\begin{document}
	
	\flushbottom
	\maketitle
	%
	%
	\thispagestyle{empty}
	
	
	
\section*{Introduction}

The concept of quantum advantage can be easily demonstrated using an example from dense coding\cite{Wilde}. 
Alice and Bob share a pair of qubits in a maximally entangled state, such as a Bell state.
By performing one-qubit unitary operations, Alice can transmit two bits of information to Bob while sending him only one physical qubit.
The information advantage of having shared access to a maximally entangled state is thus one bit in this case.
That advantage is due to an intrinsically quantum correlation in the shared resource, which in this case is entanglement\cite{Wootters}.
In this paper we discuss the realization of quantum advantage in systems that have no entanglement but possess another form of correlation: quantum discord\cite{Modern,Ollivier,Vedral}.\\%

Although entanglement has been widely used as a measure of a given state's potential quantum advantage, recent studies suggest that quantum discord is a richer resource\cite{DQC1,noentanglement1,Bobby,Discrimination,Overlap}.  
The maximum quantum advantage that can be harvested with optimal encoding is equal to the initial quantum discord. This has been shown using continuous variable and Gaussian discord\cite{Mile1}, and quantum discord for a particular discrete state\cite{Mile2}.
A recent experiment demonstrated quantum correlations in the absence of entanglement in a noisy neutron interferometer \cite{Dimitri}.
Questions remain about the relation between quantum correlations and quantum advantage, for a more general family of discrete states and non-trivial encoding schemes.
This motivates the present study of quantum advantage in systems of two qubits that are not entangled but have quantum discord, in the context of an optical experiment with bipartite $X$-states.\\

Most two-qubit and spin-1/2 states in a broad range of physical systems belong to the class of $X$-states\cite{Eberly,Rau2009}.
These include the Bell and Werner states\cite{Wilde,Werner}.
For the most general class of $X$-states, the quantum discord and the optimal projective measurement that maximize information obtained by local measurements can both be determined analytically\cite{MHR,Fanchini2010,Yichen}.
However, it is challenging to create a realistically feasible experimental system that can generate a broad range of $X$-states with arbitrary parameters, because that requires a programmable decoherence mechanism.\\

Here, we introduce an optical device that uses a single-photon source to realize general $X$-states.  
The two qubits of our single photon system are the photon polarization and the photon path in a Mach-Zehnder-like interferometer\cite{kok}.
Using only passive optical components, our system can generate mixed two-qubit $X$-states with a wide range of entanglement and quantum discord.
This device can be used for creating an $X$-state, the encoding in it of a classical random variable $K$\cite{Thomas}, and a tomographic decoding process.
We describe a series of transactions between Alice and Bob in which Alice encodes $K$ in an $X$-state and Bob attempts to decode that state.
When quantum discord is present, Bob can better estimate $K$ than he could by using only local measurements and one-qubit operations.
This quantum advantage exists even when the qubits are not entangled.
In the absence of entanglement, quantum advantage can be as large as 1/3 bit for the two-qubit system considered here.\\

Section (\ref{Theory}) discusses the theory for calculating quantum correlations and quantum advantage for a given encoding protocol.
In Sec.(\ref{experimental}) we describe an optical device for state preparation, an encoding protocol, and state tomography of symmetric $X$-states.
In Sec.(\ref{res}) we analyze the properties of the state of the system before and after the encoding, and discuss the behaviour of the quantum advantage for various encoding schemes.
We focus on the cases where there is significant quantum advantage without entanglement.

\section{Theory \label{Theory}}
\subsection{Quantum discord and two-qubit $X$-states \label{Theorya}}
Quantum discord is the difference between the total mutual information shared by two qubits, $I$, and their locally accessible (classical) mutual information $J$\cite{Modern,Wilde}. 
In light of the optical application that we propose in Sec.(\ref{experimental}), we designate the two qubits by the labels $\textrm{s}$ and $\textrm{p}$, which we use below to distinguish photon spin (polarization) from interferometric path, but which can be taken to describe any two-qubit system. 
For a system described by the density matrix $\rho _{\textrm{sp}}$ in a composed Hilbert space $ \mathcal{H}_\textrm{s}\otimes\mathcal{H}_\textrm{p}$, the total mutual information between the two qubits is :
\begin{equation}\label{mutualinf}
I(\rho _\textrm{sp})=S(\rho _\textrm{s})+S(\rho _\textrm{p})-S(\rho _\textrm{sp}),
\end{equation}%
where $S$ denotes the von Neuman entropy and $\rho _{i}=\mathrm{Tr}_{j}\left( \rho _\textrm{sp}\right) $ is the partial density matrix of the
subsystem $i$, with $i\neq j=\textrm{s},\textrm{p}$\cite{Wilde}. 
After a projective measurement\cite{POVM} on the first qubit $\Pi _{\pm}=\left|\pm \right\rangle \left\langle \pm \right|\in\mathcal{H}_\textrm{s}$, the conditional state of $\textrm{p}$ is given by $\rho _{\textrm{p}|\pm}=\mathrm{Tr}_{\textrm{s}}\left( \Pi _{\pm}\otimes\mathds{1}_{\textrm{p}}\rho _{\textrm{sp}}\right) /p_{\textrm{p}|\pm}\in\mathcal{H}_\textrm{p}$ with probability $p_{\textrm{p}|\pm}=\mathrm{Tr}\left( \Pi _{\pm}\otimes\mathds{1}_{\textrm{p}}\right)$, where $\mathds{1}_{\textrm{p}}$ is the identity for qubit $\textrm{p}$.
The classical mutual information, $J\left( \rho _{\textrm{p}|\textrm{s}}\right) $, is the amount by which the uncertainty of system $\textrm{p}$ is reduced after measuring $\textrm{s}$\cite{Modern}, and is: 
\begin{equation}
J\left( \rho _{\textrm{p}|\textrm{s}}\right) =S\left( \rho _{\textrm{p}}\right) -\inf_{\Pi _{\pm}}\sum_{\pm}S(\rho_{\textrm{p}|\pm}),
\end{equation}
where the average conditional entropy, $\sum_{\pm}S(\rho _{\textrm{p}|\pm})=p_{\textrm{p}|+}S\left( \rho _{\textrm{p}|+}\right) +p_{\textrm{p}|-}S\left( \rho _{\textrm{p}|-}\right)$, is minimized over all possible projective measurements.
Finally, quantum discord\cite{Modern}, $D\left( \rho _{\textrm{p}|\textrm{s}}\right)$, corresponds to the shared information between $\textrm{s}$ and $\textrm{p}$, that cannot be obtained by measuring $\textrm{s}$, 
\begin{equation}
D\left( \rho _{\textrm{p}|s}\right) =S(\rho _{\textrm{p}})-S(\rho _{\textrm{sp}})+\inf_{\Pi _{\pm}}\sum_{\pm}S(\rho_{\textrm{p}|\pm}).
\end{equation}
The analogous expressions for $J\left( \rho _{\textrm{s}|\textrm{p}}\right) $ and $D\left( \rho _{\textrm{s}|\textrm{p}}\right) $ can be obtained by interchanging the roles of the qubits when computing the average conditional entropy.\\

In this work we are mainly interested in symmetric two-qubit $X$-states. 
In the computational basis\cite{Wilde} an arbitrary symmetric two-qubit $X$-state takes the form:
\begin{equation}
\rho _{\textrm{sp}}=\left( 
\begin{array}{cccc}
a & 0 & 0 & w^{\ast } \\ 
0 & b & z^{\ast } & 0 \\ 
0 & z & b & 0 \\ 
w & 0 & 0 & a%
\end{array}%
\right) ,  \label{symetric}
\end{equation}%
with $b=1/2-a$. 
We show below that without loss of generality, the coherences $w\in\left[ 0,a\right] $ and $z\in\left[ 0,b\right]$ can be taken to be real and positive.
The concurrence $C(\rho_{\textrm{sp}})$ and entanglement of formation $E(\rho_{\textrm{sp}})$ are given by the Wooters formulae\cite{Wootters} 
\begin{eqnarray}
C\left( \rho _{\textrm{sp}}\right)&=&2\max \left[ 0, w -b, z -a\right], \\
E\left( C\left( \rho _{\textrm{sp}}\right)\right)&=&h\left( (1+\sqrt{1-C^{2}}%
)/2\right),
\end{eqnarray}
where $h\left( x\right) =-x\log _{2}x-\left( 1-x\right) \log_{2}\left(1-x\right) $. 
Since entanglement is a monotonic function of concurrence they have the same minima and maxima,  which are 0 and 1 respectively.\\

For such states, which are described by density matrices of the form shown in Eq.(\ref{symetric}), the conditional states after a measurement of one qubit (i.e. a local measurement) are independent of the qubit that was measured, $\rho _{\textrm{s}|\textrm{p}}=\rho _{\textrm{p}|\textrm{s}}$. 
Because of this, the average conditional entropy is also symmetric and therefore $J\left( \rho _{\textrm{s}|\textrm{p}}\right) =J\left( \rho_{\textrm{p}|\textrm{s}}\right)=J(\rho _{\textrm{sp}})$ and $D\left( \rho _{\textrm{s}|\textrm{p}}\right) =D\left( \rho_{\textrm{p}|\textrm{s}}\right) =D(\rho _{\textrm{sp}})$. 
Following the criteria introduced by Maldonado-Trapp, \emph{et al.}\cite{MHR}, the discord for this family of states is given by 
\begin{equation}  \label{discS}
D\left( \rho _{\textrm{sp}}\right) =\left\{ 
\begin{array}{ccc}
1-S\left( \rho _{\textrm{sp}}\right) +S_{\sigma _{Z}}\left( \rho _{\textrm{p}|\textrm{s}}\right) & 
\text{if} & 0\leq u\leq v, \\ 
1-S\left( \rho _{\textrm{sp}}\right) +S_{\sigma _{X}}\left( \rho _{\textrm{p}|\textrm{s}}\right) & 
\text{if} & v\leq u\leq 1,%
\end{array}%
\right.
\end{equation}
where $u=2\left( w+z\right) $ and $v=\left\vert 4a-1\right\vert $. 
The quantities $S_{\sigma _{Z}}\left( \rho _{\textrm{p}|\textrm{s}}\right) $ and $S_{\sigma _{X}}\left( \rho _{\textrm{p}|\textrm{s}}\right) $ correspond to the minimal average conditional entropies, when the optimal measurements are along $\sigma _{Z}$ and $\sigma _{X}$ respectively, where $\sigma_{i}$ are the Pauli matrices\cite{Wilde}. 
Explicitly, these entropies are given by: 
\begin{equation}  \label{SZ}
S_{\sigma _{Z}}\left( \rho _{\textrm{p}|\textrm{s}}\right) =- \frac{1-v}{2} \log _{2} \frac{1-v%
}{2} - \frac{1+v}{2} \log _{2} \frac{1+v}{2} ,
\end{equation}
and 
\begin{equation}  \label{SX}
S_{\sigma _{X}}\left( \rho _{\textrm{p}|\textrm{s}}\right) =- \frac{1-u}{2} \log _{2} \frac{1-u%
}{2} - \frac{1+u}{2} \log _{2} \frac{1+u}{2}.
\end{equation}
When $u=v$ the average conditional entropy does not depend on the measurement. 
Note that if $wz<0$ the optimal measurement $\sigma_X$ must be replaced by $\sigma_Y$\cite{MHR}. 
Although the optimal measurement changes, the value of the average conditional entropy remains the same. 
Since correlations do not change under local operations, the case $wz<0$ can be avoided by considering a rotation over the two qubits of the form 
$\exp \left(i\phi_\textrm{s}\sigma_Z/2\right)\otimes \exp \left(i\phi_\textrm{p}\sigma_Z/2\right)$. 
Thus, as noted above, we can always choose $\phi_\textrm{s}$ and $\phi_\textrm{p}$ so that $w$ and $z$ are real and positive.

\subsection{Encoding process and quantum advantage\label{encoding}}
We now consider a scenario in which two independent parties, Alice and Bob, have access to a source of discordant ($D(\rho_{\textrm{sp}})>0$) two-qubit symmetric $X$-states that are fully known to both of them. 
Alice and Bob engage in a sequence of encoding/decoding transactions, each of which starts with the same $X$-state. 
Alice generates a random variable $K$ that takes values $k=(b_1,b_2)$ with a probability distribution $p_k$, where $b_1$ and $b_2$ are random classical bits\cite{Wilde,Thomas}.
She encodes $K$ in the qubit $\textrm{s}$ and then challenges Bob to estimate $K$ by measuring the encoded state.\\
\begin{figure}[hbtp]
	\centering
	\includegraphics[width=\textwidth]{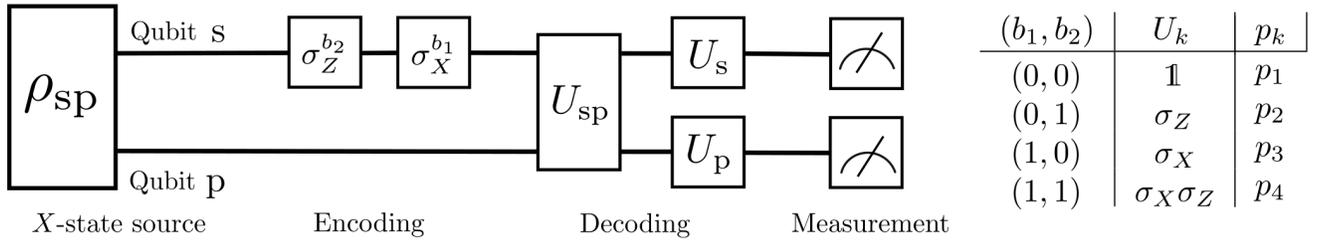}
	\caption{Quantum circuit for the encoding/decoding protocol. The upper (lower) wire
		corresponds to the qubit $\textrm{s}$ $(\textrm{p})$. Alice has access to the $X$-state source 
		$\rho_{\textrm{sp}}$. 
		She applies a local unitary operation $U_k=\sigma_{X}^{b_1}\sigma_{Z}^{b_2}$ with probability $p_k$ on
		the spin qubit s, where $b_1$ and $b_2$ are random bits. The four possible
		bit combinations are $(0,0)$, $(0,1)$, $(1,0)$ and $(1,1)$ which are related
		to the local operations $\mathds{1}_\textrm{s}$, $\sigma_Z$, $\sigma%
		_X $ and $\sigma_X\sigma_Z$ respectively. After each
		transaction Bob performs a decoding procedure consisting of two-qubit, $U_{\textrm{sp}}$ operations and local operations on polarization $U_\textrm{s}$ and path $U_\textrm{p}$.
		He then measures both qubits and estimates the value of $b_1$ and $b_2$. }
	\label{circuit}
\end{figure}

In Fig.(\ref{circuit}) we show a quantum circuit that illustrates the protocol described above.
For each transaction, Alice applies a local unitary operation $U_k=\sigma_{X}^{b_1}\sigma_{Z}^{b_2}$ to qubit $\textrm{s}$ and sends the state to Bob.
The four unitary operations that she can apply are $U_1=\mathds{1}_\textrm{s}$, $U_2=\sigma_Z$, $U_3=\sigma_X$ and $U_4=\sigma _{X}\sigma _{Z}$, where $\mathds{1}_\textrm{s}$ is the identity operator for the qubit $\textrm{s}$. 
The ensemble received by Bob is thus described by the density matrix 
\begin{equation}  \label{rhotilda}
\tilde{\rho}_{\textrm{sp}}=\sum_{k=1}^4 p_k \rho_k=\sum_{k=1}^4 p_k U_k\otimes \mathds{1}_\textrm{p}\rho_{\textrm{sp}} U_{k}^\dag\otimes \mathds{1}_\textrm{p}.
\end{equation}

By performing a decoding protocol after each transaction, Bob constructs a new random variable $K^*$.
Bob then estimates the value of $k$ that was sent by Alice, and records his estimate as $k^*=(b_1^*,b_2^*)\in K^*$. 
We will see that the accuracy of Bob's estimation is determined by his access to two-qubit operations.\\


The maximal accessible information\cite{Wilde} that Bob can obtain about $K$ corresponds to the Holevo information which for an ensemble $\epsilon =\left\{ p_{k},\rho _{k}\right\} $ is given by 
\begin{equation}
I_{q}=S(\tilde{	\rho}_{\textrm{sp}})-S(\rho _{\textrm{sp}}).  \label{Iq}
\end{equation}
Its minimum and maximum values are given by zero and two bits respectively.
When $I_q=2$ the bits $b_1$ and $b_2$ can be determined deterministically i.e Bob can recover $K$ with certainty.
An example of this case is superdense coding\cite{Wilde,Mattle}.\\

We note that when the protocol is applied to a symmetric $X$-state, the average state after the encoding $\tilde{\rho}_{\textrm{sp}}$ is also symmetric.  
Since $\tilde{\rho}_{\textrm{sp}}$ is symmetric, if Bob makes a projective measurement $\Pi _{\pm}$ on qubit $\textrm{s}$, the accessible information that he can obtain from the qubit $\textrm{p}$ is equivalent to the information that he can obtain from $\textrm{s}$ given a measurement on $\textrm{p}$, 
which is 
\begin{equation}\label{Ic}
I_{c}=\sup_{\Pi _{\pm}}\sum_{\pm}\left( S(\tilde{\rho}_{\textrm{p}|\pm})-\sum_{k}p_{k}S\left( \rho _{\textrm{p}|\pm}^{k}\right) \right) 
\end{equation}%
where $\tilde{\rho}_{\textrm{p}|\pm}=\mathrm{Tr}_{\textrm{s}}\left( \Pi _{\pm}\otimes \mathds{1}_{\textrm{p}}\tilde{\rho}_{\textrm{sp}}\right) /p_{\textrm{p}|\pm}$ and $\rho_{\textrm{p}|\pm}^{k}=\mathrm{Tr}_{\textrm{s}}\left( U_{k}^{\dag }\Pi _{\pm}U_{k}\otimes \mathds{1}_{\textrm{p}} \rho _{\textrm{sp}}\right)/p_{\textrm{p}|\pm}$ are the conditional states for qubit $\textrm{p}$. \\

Quantum advantage, $\Delta I$, is defined as the difference between $I_q$ and $I_c$, $\Delta I=I_q-I_c$\cite{Mile1,Mile2}. 
It corresponds to the extra information that Bob can gain by performing two-qubit operations prior to making local measurements of each qubit. 
When the random variable $K$ is encoded in $\rho_{\textrm{sp}}$, decoherence is induced in the system and therefore the correlations between s and $\textrm{p}$ are modified. 
The discord consumption\cite{Mile1} is defined as the difference between quantum discord before and after the encoding, $\Delta D(\rho _{\textrm{sp}})=D(\rho _{\textrm{sp}})-D(\tilde{\rho}_{\textrm{sp}})$. 
Gu et.al\cite{Mile1} proved that the quantum advantage of an encoding protocol and the discord consumption are related by the following inequality 
\begin{equation}
\Delta D(\rho _{\textrm{sp}})-J(\tilde{\rho}_{\textrm{sp}})\leq \Delta I\leq \Delta D(\rho_{\textrm{sp}}),  \label{ine}
\end{equation}%
where $J(\tilde{\rho}_{\textrm{sp}})$ is the classical mutual information between $\textrm{s}$ and $\textrm{p}$ after the protocol.
Optimal encoding is the encoding that maximizes the quantum advantage\cite{Mile1}. 
It corresponds to the one in which the total mutual information is consumed, i.e. $I(\tilde{\rho}_{\textrm{sp}})=D(\tilde{\rho}_{\textrm{sp}})=0$. 
In this case the quantum advantage is equal to the initial amount of quantum discord, $\Delta I=D(\rho_{\textrm{sp}})$.

\section{Experimental Proposal\label{experimental}}
\begin{figure}[tbh]
\centering
\includegraphics[width=\textwidth]{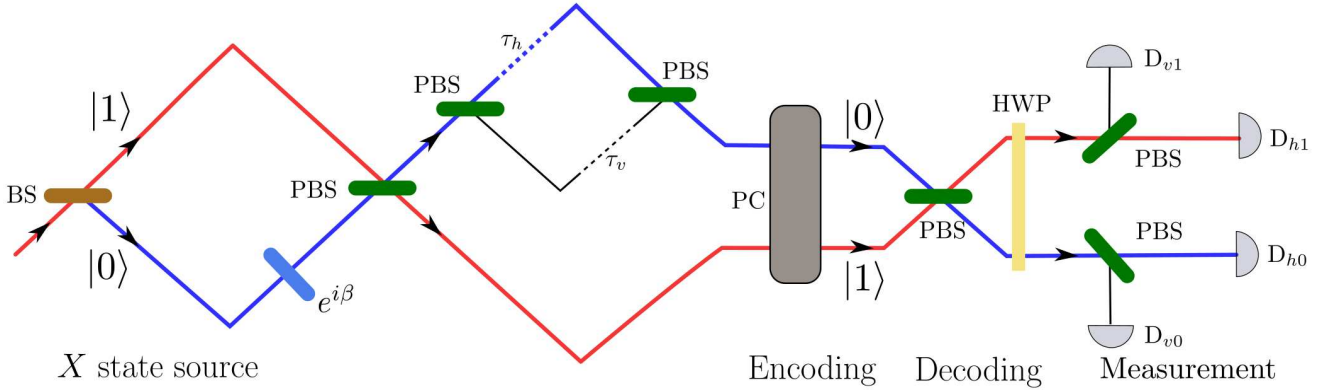}
\caption{ Theoretical proposal for preparation of an $X$-state and encoding and decoding of a random variable. 
\textbf{X-state source}: A polarized photon in an equal superposition of vertical and horizontal component, $\frac{\left\vert h\right\rangle +\left\vert v\right\rangle }{\protect\sqrt{2}}$, is incident on a beam splitter (BS), with reflection coefficient $R$ and transmission coefficient $T$. 
In path $\left\vert 1\right\rangle $ the photon experiences a random phase shift, $\beta $, with Gaussian probability distribution to make it incoherent with $\left\vert 0\right\rangle $. 
Paths $\left\vert 0\right\rangle $ and $\left\vert 1\right\rangle $ are recombined with a polarizing beam splitter (PBS) that transmits $\left\vert h\right\rangle $ and reflects $\left\vert v\right\rangle $. 
In path $\left\vert 0\right\rangle $ the photon goes through two PBSs. 
This adds time delays $\tau_{h}$ and $\tau_{v}$ for the horizontal and vertical components of path $\left\vert 0\right\rangle $ respectively. 
\textbf{Encoding}: Both paths $\left\vert 0\right\rangle $ and $\left\vert 1\right\rangle $ go through a polarization controller (PC).
The PC can arbitrarily rotate the polarization state of a photon.
\textbf{Decoding}: Paths paths $\left\vert 0\right\rangle $ and $\left\vert 1\right\rangle$ are recombined at PBS.
Both paths go through a half wave plate (HWP) at an angle $\pi/8$ with respect to the horizontal.
\textbf{Measurement}: Two PBSs  and four photon detectors are used to measure in the $\{\left\vert h\right\rangle \left\vert 0\right\rangle ,\left\vert h\right\rangle \left\vert 1\right\rangle ,\left\vert v\right\rangle \left\vert 0\right\rangle ,\left\vert v\right\rangle \left\vert 1\right\rangle \}$ basis.}
\label{setup}
\end{figure}

In this section we propose an optical implementation for generating symmetric two-qubit $X$-states, as in Eq.(\ref{symetric}), and use them in an encoding protocol. The detailed setup, shown in Fig.(\ref{setup}), consists of three parts: an $X$-state source, an encoding and decoding mechanism, and a measurement process. \\

We use a linear polarization basis with horizontal and vertical components designated by $\left| h\right\rangle $ and $\left| v\right\rangle $ respectively, and a path basis designed by  $\left|0\right\rangle $ and $\left| 1\right\rangle $, which in the computational basis correspond to 
\begin{equation}
\left\vert h\right\rangle =\left( 
\begin{array}{c}
1 \\ 
0%
\end{array}%
\right) \text{, }\qquad\left\vert v\right\rangle =\left( 
\begin{array}{c}
0 \\ 
1%
\end{array}%
\right)
\text{, }\qquad\left\vert 0\right\rangle =\left( 
\begin{array}{c}
1 \\ 
0%
\end{array}%
\right)\text{, }\qquad\left\vert 1\right\rangle =\left( 
\begin{array}{c}
0 \\ 
1%
\end{array}%
\right)
\end{equation}

In this basis, an $X$-state can be thought as an incoherent superposition of two Bell-like states, $\left| \Phi \right\rangle = c_{h0}\left| h \right\rangle\left|  0\right\rangle +c_{v1}\left| v \right\rangle\left| 1\right\rangle$ and $\left| \Psi \right\rangle = c_{h1}\left| h \right\rangle\left|  1\right\rangle +c_{v0}\left| v \right\rangle\left|  0\right\rangle$, where the coefficients $c_{ij}$ are the probability amplitudes of the states $\left| i \right\rangle\left|  j\right\rangle$. 
It is well known in optics that a polarizing beam splitter (PBS) can create such states for a single incoming photon by properly choosing the PBS input port. 
Then, by combining the two incoherent paths at the input ports of a PBS the output state will be an $X$-states of the form (\ref{symetric}).\\

For simplicity, our $X$-state source uses an input photon with a polarization state $ \left\vert \psi _{\mathrm{in}}\right\rangle =\left( \left\vert
h\right\rangle +\left\vert v\right\rangle \right) /\sqrt{2}$. 
The state is then split in two paths and a random phase is added to one path to make them relatively incoherent. 
This can be done by a random switch choosing which path the photon goes through, or by a beam splitter (BS) and a subsequent path delay longer than the coherence length of the photon, or by adding random noise in one of the paths. 
In particular, we will describe this process by assuming that the input photon first encounters a beam splitter BS with transmission and reflection coefficients $T $ and $R$ respectively, where $T+R=1$, and introducing a random source of phase noise in one path, see Fig.(\ref{setup}). 
It provides a phase shift $\exp \left( i\beta \right) $ where $\beta $ is a random variable with a Gaussian probability distribution $\exp \left( -\beta ^{2}/\left( 2\sigma ^{2}\right) \right) /\sqrt{2\pi \sigma ^{2}}$ and standard deviation $\sigma $. 
We consider $\sigma $ to be sufficiently large such that the average of different phases $\left\langle \exp\left(\pm i\beta \right)\right\rangle =\exp\left(-\frac{\sigma ^{2}}{2}\right)$ can be neglected, producing two incoherent beams.
Paths $\left\vert 0\right\rangle $ and $\left\vert 1\right\rangle $ are recombined with a polarizing beam splitter PBS; all polarizing beam splitters in this apparatus transmit polarization $\left\vert h\right\rangle $ and reflect polarization $\left\vert v\right\rangle $. 
The PBS acts like a control not (CNOT) gate with polarization and path as control and target qubits respectively\cite{Wilde}. 
In path $\left\vert 0\right\rangle $ we place a PBS that introduces one auxiliary path (black path in Fig.(\ref{setup}) ). 
In those paths we add a controllable time delay, $\tau _{h}$ or $\tau _{v}$, which allows us to control the coherences of the $\left\vert h\right\rangle $ and $\left\vert v\right\rangle $ components of $\left\vert 0\right\rangle $. 
With this, the anti-diagonal term associated with $\{\left\vert h\right\rangle \left\vert 0\right\rangle \}$ and $\{\left\vert v\right\rangle \left\vert 0\right\rangle \}$ decreases by a factor of $\kappa _{h}=e^{-\frac{\tau _{h}}{\tau _{c}}}$ and $\kappa _{v}=e^{-\frac{\tau _{v}}{\tau _{c}}}$ respectively, where $\tau_c$ is the coherence time of the photons.  
The resulting density matrix can be written in the  $\{\left\vert h\right\rangle \left\vert 0\right\rangle ,\left\vert h\right\rangle
\left\vert 1\right\rangle ,\left\vert v\right\rangle \left\vert 0\right\rangle ,\left\vert v\right\rangle \left\vert 1\right\rangle \}$ basis as 
\begin{equation}
\rho =\frac{1}{2}\left( 
\begin{array}{cccc}
R & 0 & 0 & -iR\kappa _{h} \\ 
0 & T & -iT\kappa _{v} & 0 \\ 
0 & iT\kappa _{v} & T & 0 \\ 
iR\kappa _{h} & 0 & 0 & R%
\end{array}%
\right).  \label{rhooo}
\end{equation}

To encode the random variable $K$, both paths $\left\vert 0\right\rangle $ and $\left\vert 1\right\rangle $ go  through a polarization controller, PC.
The PC acts simultaneously in both paths and allows Alice to arbitrarily rotate the polarization state of the photon.
We assume that Alice has a random number generator that tells her which set of bits $\left( b_{1},b_{2}\right) $ to send in each transaction.
While to send the bits $(0,0)$ she does nothing to the polarization state, to send $(0,1)$, $(1,0)$ or $(1,1)$, she applies a rotation at $\pi$ around the axis $z$, $x$ and $y$ in the polarization Bloch sphere\cite{Wilde}. 
The decoding process is performed by a PBS and half wave plate (HWP) which acts as a CNOT and a Hadamard gate in polarization respectively\cite{Wilde}. 
The HWP is set at an angle $\pi/8$ with respect to the horizontal and transforms the polarization components as $\left\vert h\right\rangle\to (\left\vert h\right\rangle+\left\vert v\right\rangle)/\sqrt{2}$ and $\left\vert v\right\rangle\to (\left\vert h\right\rangle-\left\vert v\right\rangle)/\sqrt{2}$.
A PBS in each path $\left\vert 0\right\rangle $ and $\left\vert 1\right\rangle $ are used to measure in the $\{\left\vert h\right\rangle \left\vert 0\right\rangle ,\left\vert h\right\rangle
\left\vert 1\right\rangle ,\left\vert v\right\rangle \left\vert 0\right\rangle ,\left\vert v\right\rangle \left\vert 1\right\rangle \}$ basis with detectors D$_{h0}$, D$_{h1}$, D$_{v0}$ and D$_{v1}$.

\section{Results \label{res}}
\subsection{Quantum correlations before encoding\label{before}}
Since the von Neuman entropy is invariant under local operations, we apply to the density matrix of the experimental $X$ states the following rotational operation, $e^{i\frac{\pi }{4}\sigma _{Z}}\otimes I_{\textrm{p}}\rho e^{-i\frac{\pi }{4}\sigma _{Z}}\otimes I_{\textrm{p}}$, such that the density matrix elements of the rotated state are real and positive.
The correlations present in the original, $\rho$, and rotated state are the same and they can be computed using the formulas discussed in Sec (1.1). 
In particular, concurrence is given by $C\left( \rho \right) =\max \left[ 0,\kappa _{h}R-T,\kappa _{v}T-R\right] $ and it vanishes when $\kappa _{h}R\leq T$ and $\kappa _{v}T\leq R$, or when $R=T$. 
The analytical expression for quantum discord is given by equation (\ref{discS}), with $u=\kappa _{h}R+\kappa _{v}T$ and $v=\left\vert 2R-1\right\vert $. 
By using the second derivative test\cite{Thomas} we found that, with respect to variations of $\kappa _{h}$ and $\kappa _{v}$,  quantum discord is maximized in two cases: $\kappa _{v}=1$ and $\kappa _{h}=0$; and $\kappa _{v}=0$ and $\kappa _{h}=1$.\\
\begin{figure}[tbph]
	\centering
	\includegraphics[width=\textwidth]{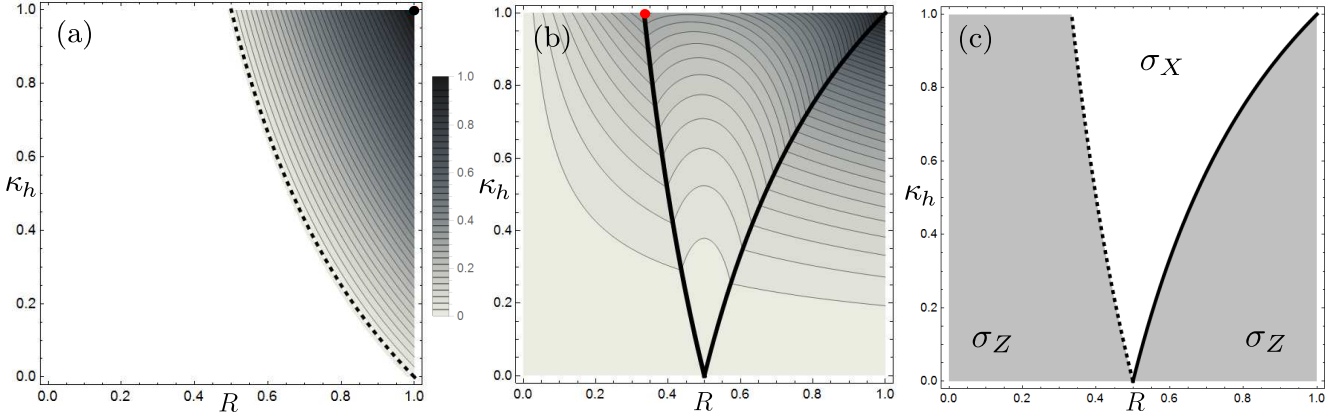}
	\caption{Concurrence, $C(\rho )$, (a) and quantum discord, $D(\rho )$, (b) as a function of $R$ and $\kappa_h$, for $\kappa_v=0$. 
	(c) Parameter regions of the optimal measurements.
	(a) the dashed line indicates where the concurrence becomes zero.
	The red point corresponds to the Bell state $\left|\psi_{+} \right\rangle$ for which concurrence is maximum.
	(b) the red dot corresponds to the maximum value of discord that can be reached without concurrence.
	The gray and white regions of Fig. (c) denote where $\sigma_Z$ and $\sigma_X$ are the measurement that minimizes the average conditional entropy, respectively.
	The states along the solid ($R=1/(2-\kappa_h)$) and dashed ($R=1/(2+\kappa_h)$) boundaries correspond to the Werner and Werner-like states respectively.}
	\label{fig:before}
\end{figure}

Figure (\ref{fig:before}) (a) and (b) show the behaviour of the pre-encoding concurrence and discord, $C\left( \rho \right) $ and $D\left( \rho \right) $, as a function of $R$ and $\kappa _{h}$ when $\kappa _{v}=0$. 
When $\kappa _{h}=0$, the corresponding dependence on $R$ and $\kappa _{v}$ can be found by reflecting these figures about the line $R=1/2$. 
From Fig.(\ref{fig:before})(a) we note concurrence increases monotonically with $\kappa_h$ and decreases monotonically with $R$. 
Concurrence is zero at the region left of the dashed line, $(1+\kappa_h)R=1$. 
It reaches its maximum value, $C=1$, when the photon is completely reflected onto path $\left\vert 0\right\rangle $, $R=1$, and there is no time delay in the auxiliary path, $\kappa _{h}=1$.
This maximum corresponds to the Bell state indicated by the red point, $\left|\psi_{+} \right\rangle=\left( \left|h0 \right\rangle +\left|v1 \right\rangle\right)/\sqrt{2}$\cite{Wilde}.
Fig.(\ref{fig:before})(b) shows that quantum discord also increases monotonically with $\kappa _{h}$. 
Discord only vanishes when the photon is completely transmitted, $R=0$, or when the coherence time of the photon is much less than the time delay, $\kappa _{h}=0$.
It also reaches its maximum value for Bell states, $D=1$, at $R=1$ and $\kappa _{h}=1$.
Besides the global maximum, discord also has local maxima along the black boundaries.
The highest value that discord reaches without entanglement is represented by the red dot and its value is $D=1/3$ at $\kappa _{h}=1$ and $R=1/3$.
In Fig.(\ref{fig:before})(c) the gray and white regions show the regions where the optimal measurements are $\sigma _{Z}$ and $\sigma _{X} $ respectively. 
These are the measurements of polarization that minimize the average conditional entropy and therefore they least disturb the system but allow more information about the system to be obtained.
The conditional entropy is independent of the measurement for the states along the black lines.
The states along the black solid line are the Werner states\cite{Wilde,Werner} which can be written as $\rho_W=(1-R\kappa_h)\mathds{1}_{\textrm{sp}}/4+R\kappa_h\left|\psi_{+} \right\rangle\left\langle\psi_{+}\right|$.
We indicate the Werner-like states to the states by the black dashed line, which can be written as $(1+R\kappa_h)\mathds{1}_{\textrm{sp}}/4-R\kappa_h\left|\psi_{+} \right\rangle\left\langle\psi_{+} \right|$.

\subsection{Quantum correlations after encoding\label{after}}
To maximize quantum correlations we consider the pre-encoding state as the one that satisfies $\kappa _{h}=1$ and $\kappa _{v}=0$, this is
\begin{equation}
\rho _{\textrm{sp}}=\frac{1}{2}\left( 
\begin{array}{cccc}
R & 0 & 0 & R \\ 
0 & T & 0 & 0 \\ 
0 & 0 & T & 0 \\ 
R & 0 & 0 & R%
\end{array}%
\allowbreak \right).  \label{rhosp}
\end{equation}%
After each transaction of the pair of bits $(b_1,b_2)$ according to the encoding protocol described in Sec.(\ref{encoding}), the density matrix of the post-encoding state becomes
\begin{equation}
{\small \rho _{k}=\frac{1}{4}\left( 
	\begin{array}{cccc}
	1-\left( -1\right) ^{b_{1}}\left( T-R\right)  & 0 & 0 & \left( -1\right)
	^{b_{2}}\left( 1+\left( -1\right) ^{b_{1}}\right)R  \\ 
	0 & 1+\left( -1\right) ^{b_{1}}\left( T-R\right)  & \left( -1\right)
	^{b_{2}}\left( 1-\left( -1\right) ^{b_{1}}\right)R  & 0 \\ 
	0 & \left( -1\right) ^{b_{2}}\left( 1-\left( -1\right) ^{b_{1}}\right)R  & 
	1+\left( -1\right) ^{b_{1}}\left( T-R\right)  & 0 \\ 
	\left( -1\right) ^{b_{2}}\left( 1+\left( -1\right) ^{b_{1}}\right)R  & 0 & 0
	& 1-\left( -1\right) ^{b_{1}}\left( T-R\right) 
	\end{array}%
	\right).\label{b1b2}}
\end{equation}%
From Eq.(\ref{b1b2}) we immediately note that $b_2$ only appears in the non-diagonal terms.
Without joint measurements Bob cannot determine the value of $b_2$, thus he will not be able to estimate $K$ with certainty.\\
\begin{figure}[tbph]
	\centering
	\includegraphics[width=\textwidth]{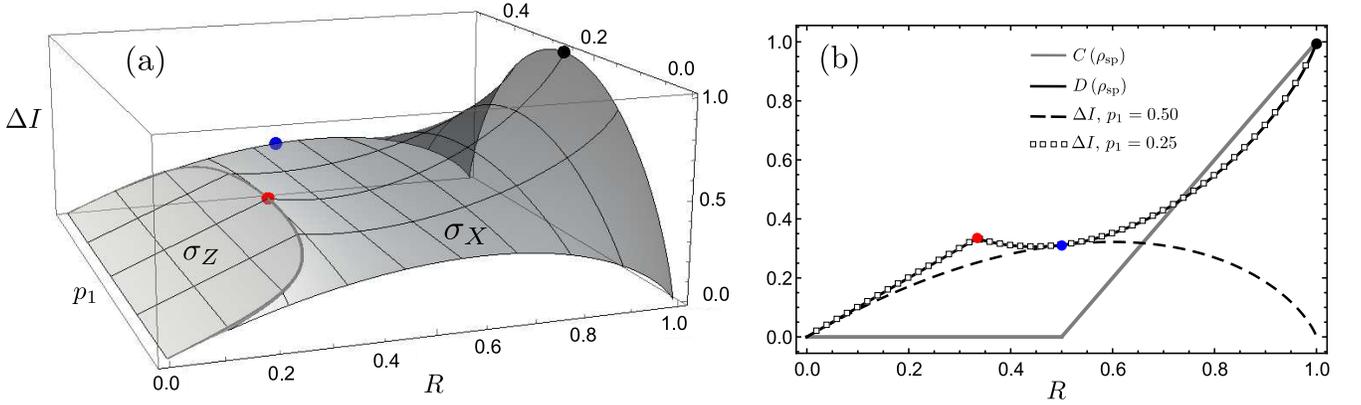}
	\caption{(a) Quantum advantage, $\Delta I$,  as a function of $R$ and $p_1$ for the quasi-optimal encoding. 
			(b) Pre-encoding $C(\rho_{\textrm{sp}})$ (gray line); pre-encoding discord $D(\rho_{\textrm{sp}})$ (black squares); quantum 					advantage $\Delta I$ with $p_1=0.25$ (black line); $\Delta I$ with $p_1=0.5$ (dashed line).
			In (a) the gray boundary denotes where the optimal measurement for $I_c$ changes from $\sigma_Z$ to $\sigma_X$.
			In (a) and (b) the black point denotes the Bell state $\left|\psi_{+} \right\rangle$ that maximizes the value of $\Delta I$ when 						$p_1=0.25$. 
			The red and blue points denote the maximum value of $\Delta I$ when there is no entanglement for $p_1=0.25$ and $p_1=0.5$ respectively.}
	\label{fig:after}
\end{figure}

After averaging out a series of transactions, on average Bob receives the state 
\begin{equation}
{\small \tilde{\rho}_{\textrm{sp}}=\frac{1}{2}\left( 
	\begin{array}{cccc}
	(p_{3}+p_{4})T+(p_{1}+p_{2})R & 0 & 0 & (p_{1}-p_{2})R \\ 
	0 & (p_{3}+p_{4})R+(p_{1}+p_{2})T & (p_{3}-p_{4})R & 0 \\ 
	0 & (p_{3}-p_{4})R & (p_{3}+p_{4})R+(p_{1}+p_{2})T & 0 \\ 
	(p_{1}-p_{2})R & 0 & 0 & (p_{3}+p_{4})T+(p_{1}+p_{2})R%
	\end{array}%
	\right) ,}  \label{rhotildai}
\end{equation}%
where $p_{1}$, $p_{2}$, $p_{3}$ and $p_{4}$ are the probabilities described in section (\ref{encoding}). 
Without losing generality we consider only positive coherences $p_{1}\geq p_{2}$ and $p_{3}\geq p_{4}$.
For this ensemble, concurrence is given by 
\begin{equation}
C\left( \tilde{\rho}_{\textrm{sp}}\right) =\max \left[0,\left( 2p_{1}-1\right) R-(p_{1}+p_{2})T,\left( 2p_{3}-1\right)R-(p_{3}+p_{4})T\right].
\end{equation}
Quantum discord is given by equation (\ref{discS}), with $u=\left( 2\left( p_{1}+p_{3}\right) -1\right) R$ and $v=\left\vert \left( p_{3}+p_{4}\right) \left( 1-2R\right) +R\right\vert $.

\subsection{Quantum advantage and quasi-optimal encoding \label{advantage}}
As noted in Sec.(\ref{encoding}), an optimal encoding is the one that consumes all correlations, $I(\tilde{\rho}_{\textrm{sp}})=0$\cite{Mile1}.
This is satisfied when the final average state is $\frac{1}{4}\mathds{1}_{\textrm{s}}\otimes \mathds{1}_{\textrm{p}}$, in other words it is the encoding that introduces the greatest amount of decoherence to the system.
From Eq.(\ref{rhotildai}) we note that an optimal encoding must satisfy one of the following conditions: $p_{1}=p_{2}=p_3=p_4=1/4$, or $R=T$.
It may seem counter-intuitive that the encoding that maximizes the quantum advantage, $\Delta I=I_q-I_c$, is the one that leads to the maximally mixed state, however, the identity is the result of the average of a large number of transactions, and the quantum advantage depends on the non-local information in each one of them.\\

Aside from the noted relationship\cite{Mile1} between initial quantum discord and the quantum advantage in an optimal encoding, $\Delta I=D(\rho _{\textrm{sp}})$, we found that in this case there is a direct relationship between the initial amount of information and the quantities $I_c$ and $I_q$.
The total accessible information, $I_q$, can be conceived as the amount of mutual information that has been removed from the initial state, $I_{q}=I(\rho_{\textrm{sp}})-I(\tilde{\rho}_{\textrm{sp}})$.
Since for optimal encodings the total mutual information becomes zero, we can state that these are the encodings that extract all the information from the system and therefore $I_q=I(\rho _{\textrm{sp}})$.
In addition to this, from  Eq.(\ref{Iq}) we note that $I_q$ is also related to the randomness that has been introduced to the system. 
For example, when only one of the events $k$ has preference, say $p_2=1$, no randomness is introduced and $I_q$ is zero, while in an optimal encoding all the events have the same probability, $p_k=1/4$, and no more randomness can be introduced.
Since for optimal encodings $\Delta I=D(\rho _{\textrm{sp}})$ and $I_q=I(\rho_{\textrm{sp}})$ it is clear that the locally accessible information coincides with the initial classical information, $I_{c}=J(\rho _{\textrm{sp}})$.\\

Although the optimal encoding is the one that maximizes the quantum advantage it is also interesting to study a more general encoding that we call quasi-optimal, in which $p_1=p_2$ and $p_3=p_4$.
For this encoding all quantum discord is consumed, $D(\tilde{\rho}_{\textrm{sp}})=0$, but there is a remainder of classical mutual information, $I(\tilde{\rho}_{\textrm{sp}})=J(\tilde{\rho}_{\textrm{sp}})$.  
This allows us to study quantum correlations and advantage for a range of $p_1$, specifically $p_1\in\left[0,1/2 \right]$.\\

Fig.(\ref{fig:after})(a) shows quantum advantage, $\Delta I$, for the quasi-optimal encoding as a function of $R$ and  $p_1$.
The maximum advantage that Bob can obtain with non-zero entanglement is $\Delta I=1$, which is reached at $p_1=1/4$ and $R=1$.
This corresponds to the Bell state $\left|\psi_{+} \right\rangle$ (black dot). 
The maximum value of $\Delta I$ that can be reached without entanglement (red dot) is $\Delta I=1/3$ at $p_1=1/4$ and $R=1/3$.
Quantum advantage only vanishes when $R=0$, this is when the photon is completely transmitted by the variable BS. 
The complete transmission creates a Bell state after the PBS that becomes completely incoherent by making $\kappa_v=0$. 
This is why it cannot be made more random in the encoding process. 
The gray boundary denotes where the measurement that optimizes $I_c$ changes from $\sigma_z$ to $\sigma_x$.
This measurement corresponds to the one that Bob must perform in order to obtain the most information when he is restricted to one-qubit operations.
In Fig.\ref{fig:after}(b) we compare quantum advantage as a function of $R$ for $p_1=1/2$ (dashed line) and $p_1=1/4$ (black line) concurrence (grey line) and quantum discord (black squares). For optimal encoding, i.e. $p_1=1/4$, the quantum discord and the quantum advantage is exactly the same as we noted before. 
For $p_1=1/4$, the global maximum of the quantum advantage is 1 bit at $R=1$, as indicated by the black dot. 
This corresponds to the Bell state $\left|\psi_{+} \right\rangle$.
In the region where there is no entanglement, the maximum quantum advantage is $1/3$ bit at $R=1/3$, as indicated by the red dot. 
For  $p_2=1/2$, the quantum discord and the quantum advantage are in general different. 
These two values become the same at around $R=1/2$, where the quantum advantage for $p_2=1/2$ reaches the maximum of $\Delta I=3(2-\log_23)/4\approx0.311$ bit, as indicated by the blue dot.
Although the maximum value of the quantum advantage without entanglement is $\Delta I\approx0.333$, the case denoted by the blue dot is useful since it corresponds to the encoding scheme where  Alice only needs to apply $\sigma_X$ whereas in the optimal encoding she has to apply the three Pauli matrices, $\sigma_X$, $\sigma_Y$ and $\sigma_Z$. For this case the quantum advantage is significant at  $\Delta I\approx0.311$, and is much easier to perform the protocol in a experimental implementation.  

\section{Conclusions}

In this work, we proposed a versatile optical realization of symmetric two-qubit $X$-states. 
We discussed the quantum correlations in these states and the quantum advantage that be realized in various encoding and decoding schemes.
Our approach seems experimentally accessible. Its simplicity facilitates analyses of quantum discord and its role in attaining quantum advantage. 
We show explicitly that significant quantum advantage can be attained even when there is no entanglement. 
We prove for the first time that $1/3$ bit is the maximum value of quantum advantage that can be attained in symmetric two-qubit $X$-states without entanglement. 
We also find that significant quantum advantage can be attained with simplified encoding/decoding protocols. 
A protocol with only one local unitary operation can achieve quantum advantage of $0.311$ bit, which is 93\% of the maximum value. 
Our work demonstrates the significance of quantum discord in determining quantum advantage in encoding/decoding protocols, and suggests that quantum discord is superior to entanglement as an estimator of quantum advantage.

\section*{Acknowledgements (not compulsory)}
The research is supported by the Chilean National Commission for Scientific and Technological Research, CONICyT (A.M-T.) and the American National Science Foundation, NSF 1307416 and 1430094), (P.S.).

\section*{Author contributions statement}
A.M-T. and P.S. conceived the idea. A.M-T. and A.H. formalized the theory. P.S. designed the experimental proposal. C.C. guided the work. A.M-T. wrote the manuscript, with contributions from the other authors.  All authors discussed the results and commented on the manuscript at all stages.
\end{document}